# Confinement-induced Ultrafast Conductivity in 2D Perovskites resolved by correlative Terahertz-NIR Spectroscopy

**Lion Krüger[1#], Fabian Brütting[1,4#], Michael Baumann[2], Moritz B. Heindl[1], Maximilian Spies[3], Anna Köhler[3], Alexander JC Kühne[2], Georg Herink[1*]**

[1] Experimental Physics VIII - Ultrafast Dynamics, University of Bayreuth, Bayreuth, Germany
[2] Institute of Organic and Macromolecular Chemistry, Ulm University, Ulm, Germany
[3] Soft matter Optoelectronics, Bayreuth Institute of Macromolecular Research (BIMF) and Bavarian Polymer Institute (BPI), University of Bayreuth, Bayreuth, Germany
[4] School of Chemistry, The University of Melbourne, Parkville, Victoria, Australia
[#] Co-first authors of this article
* Author to whom correspondence should be addressed: georg.herink@uni-bayreuth.de

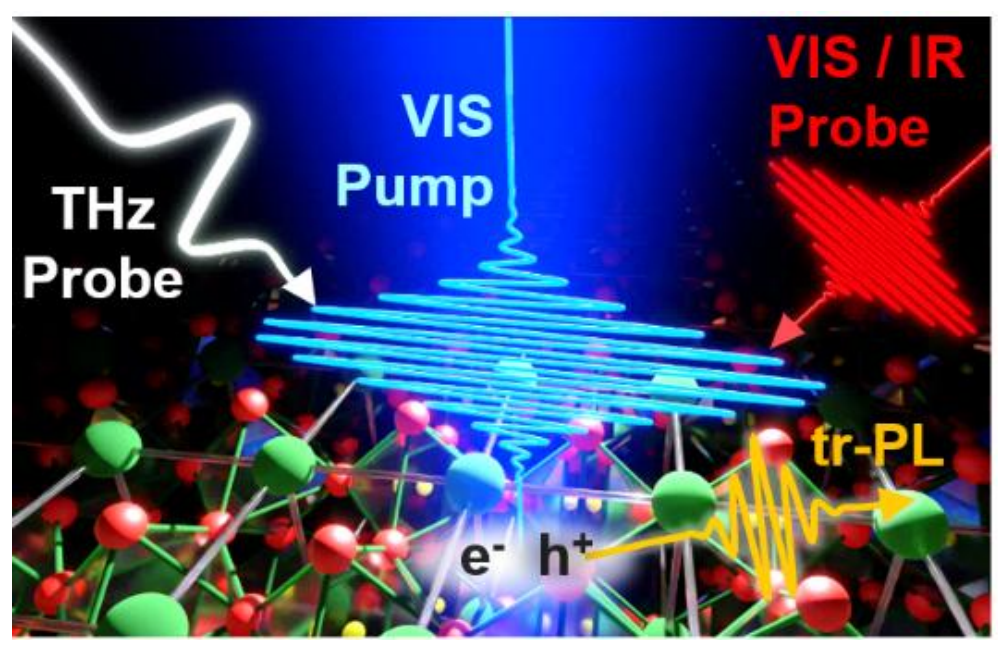

**Abstract:** Quantum wells made of quasi two-dimensional organic-inorganic hybrid perovskites (2D-PKs) offer a high degree of flexibility in tailoring optoelectronic properties through carrier confinement and functional interlayers. Compared to their 3D counterparts, 2D-PKs exhibit tunable photoluminescence, excitonic binding at room temperature and enhanced structural stability. However, the dynamics of photo-induced charge carriers and their transport properties are highly intertwined due to the interplay of diverse excitation species, charge carrier cooling, transport, and radiative and non-radiative recombination. In this study, we employ optical-pump terahertz-probe spectroscopy (OPTP) to analyze the local conductivity dynamics of 2D-PK (n = 5) 2-(9H-carbazol-9-yl)ethan-1-ammonium ($(MA)_5(CzEA)_2(PbI_3)_5$) and 3D-MAPI methylammonium lead iodide ($MeNH_3PbI_3$) perovskites at timescales down to picoseconds. Remarkably, we observe an intensity-dependent, 2D-specific buildup of an ultrafast, few-picosecond decay in local THz-conductivity. By combining OPTP with transient absorption (TA) and picosecond time-resolved photoluminescence (TRPL), we correlate photoconductivity and carrier population. Thus, can attribute the 2D-specific ultrafast THz response to delayed hot-carrier cooling and subsequent exciton formation, which effectively reduces the free-carrier conductivity. This intensity-dependent, ultrafast THz response appears as a signature of the recently identified hot-carrier bottleneck in bulk perovskites, and this effect manifests itself in a unique form in the 2D material. These results encourage further investigations on the impact of functional organic interlayers and provide insights into designing tunable carrier responses for ultrafast devices via adapted heterostructures and confinements.

## I. INTRODUCTION

2D hybrid organic-inorganic perovskites (2D-PKs) represent an exciting alternative to classical inorganic materials, since they combine broadly tuneable optoelectronic semiconductor properties of hybrid perovskites with additional degree of freedom via highly anisotropic material compositions as well as dielectric and quantum confinement. 2D-PK quantum wells are solution processable and the stoichiometry between small methyl ammonium- (MA) and large organic ammonium-ions controls the active layer thickness *n*. The organic interlayers induce dielectric confinement, i.e. reduce Coulomb screening of electrons and holes in the PK semiconductor and stabilize excitons at room temperature[1]. In addition, 2D-PKs exhibit superior resistance against ion migration and humidity compared to the 3D counterparts. Accordingly, 2D-PKs represent an attractive material system for next-generation photonic applications in micro-LEDs and lasers[2] and also as a protective additive in solution-processable solar cells[3,4].

Due to the complex photophysical response, insights into the primary steps of light-matter interaction and the nature of charge transport are highly intertwined. The intrinsic material features diverse charge carrier recombination channels, including debated hot carrier effects, exciton formation and transport[5]. On the other hand, the added structural and compositional degrees of freedom introduce novel degrees to engineer carrier dynamics[6], and also render the material susceptible to inhomogeneities and varying degrees of confinements within the sample.

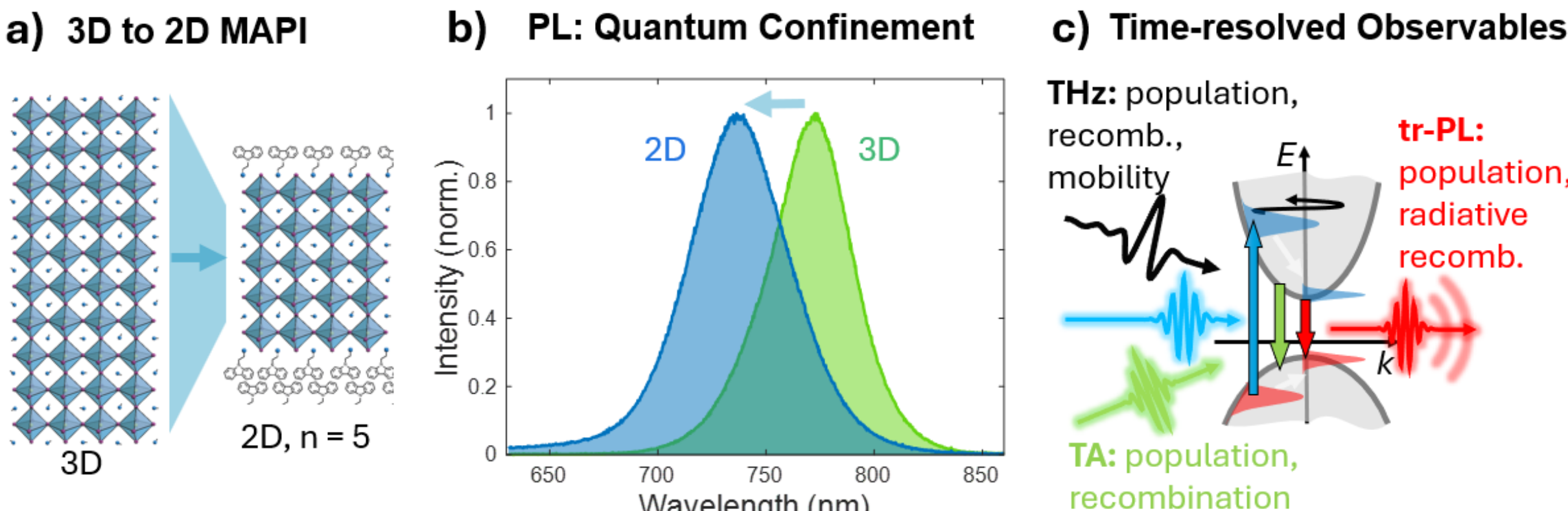

**FIG 1:** a) The impact of confinement from 3D-MAPI to 2D quantum wells with layer thickness n = 5 is studied. b) Photoluminescence spectra (PL) display a blue-shift due to the quantum confinement in 2D-PK. c) This study correlates a set of time-resolved observables to access population and mobility dynamics: optical-pump THz-probe spectroscopy (OPTP), transient absorption (TA) and time-resolved photoluminescence (tr-PL).

Time-resolved Terahertz (THz) spectroscopy in combination with transient absorption and photoluminescence spectroscopy has recently been established for non-contact access to initial photo-induced carrier and conductivity dynamics in hybrid perovskite materials[7–11]. Previous transient absorption studies could resolve a remarkably slow, intensity-dependent cooling of hot carriers for above-bandgap excitation compared to conventional semiconductors[12,13]. The effect has been attributed to a “hot phonon” bottleneck (HPB) with the buildup of a hot phonon population and the saturation of efficient carrier cooling via reduced phonon-phonon scattering[12,14]. The dynamics have been also related to polaron formation in the soft ionic lattice[15] and reduced electron-phonon coupling due to potential screening of the underlying Fröhlich interaction[12].

The impact of confinement on these complex charge carrier dynamics in the emerging class quantum-confined 2D-perovskites currently poses fundamental and unresolved questions. Interestingly, recent studies report two cooling regimes and an enhanced phonon bottleneck effect[16–19], but also the absence of a phonon bottleneck effect for n = 1 2D-PKs[20,21] – pointing towards a sensitive dependence of carrier relaxation on the specific 2D material and its n-value.

Also, the Quantum confinement is expected to enhances early bimolecular and Auger recombination channels[22]. Furthermore, THz-studies report a reduced trap-related recombination[22] and moderate to high mobilities for the anisotropic 2D sheet conductivity with respect to the 3D material[23,24]. Yet, interactions with the organic interlayers add further complexity to the photophysical response but thereby open-up diverse avenues to modify the material response[24] and design future optoelectronic function in well-controlled morphologies.

In this work, we investigate the impact of structural confinement onto photoinduced charge carrier dynamics and mobility for prototypical 2D-PK thin films against the 3D material, as illustrated in Figure 1. We probe the material systems with correlated time-resolved THz, transient absorption and ultrafast photoluminescence spectroscopy to gain access to the dynamics of photogenerated charge carriers and local conductivity.

## II. EXPERIMENTAL METHODS

### A. Ultrafast time-resolved THz/NIR/VIS spectroscopy

We access the evolution of carrier population, local conductivity and luminescence emission by combining a set of correlative time-resolved ultrafast spectroscopies, including optical-pump Terahertz-probe spectroscopy (OPTP), visible/near-infrared transient absorption (TA) and picosecond temporally-resolved photoluminescence (tr-PL).

Optical-pump Terahertz-probe spectroscopy (OPTP) is based on an Ytterbium-based femtosecond amplifier system (Light conversion, Pharos) with 150-fs pulses at an energy of 1 mJ and repetition rate of 10 kHz. The generation of single-cycle THz pulses at 1 THz is based on optical rectification in $LiNbO_3$ in the tilted-pulse front phase-matching geometry, as described in Ref. ([25]). Second harmonic generation (SHG) of a fraction of the laser fundamental generates visible pump pulses at 515 nm. Pump and THz pulses are co-focused onto the sample. THz-pulses are detected via electro-optical sampling employing the Pockels effect in a 1 mm-GaP crystal. Sampling pulses are generated by an optical-parametric amplifier (OPA) that is pumped with a small fraction of the laser amplifier. The pump pulses are chopped for sensitive digital lock-in detection.

Transient absorption spectroscopy in the visible/near-infrared spectrum is implemented via the SHG-based pump pulses and tunable probing pulses from the OPA. The probe wavelength is tuned to the minimum of the conduction band / exciton transition of 750 nm and 720 nm of the 3D/2D PK, respectively.

Time-resolved photoluminescence spectra (tr-PL) have been acquired with a streak-camera / spectrograph combination (Hamamatsu C5680, Synchro-Scan) with a temporal resolution down to < 10 ps. The samples have been excited with SHG ($\lambda = 415$ nm) of an actively-modelocked Ti:Sapphire oscillator with a repetition rate of 80 MHz.

### B. Material synthesis

The synthetic details of the 2D quantum wells are described in the Supporting Information. In short, we prepare stoichiometric solutions of the respective perovskite precursors (n = 5, n = ∞) in a mixture of DMF and DMSO (1:1). The solutions are spin coated on quartz substrates, where perovskite formation is induced by applying chlorobenzene as an antisolvent. The obtained films are baked at 100 °C and then exposed to methyl amine gas to obtain 377±47 nm thick 2D films. The final films are encapsulated with a layer of PMMA.

To prepare the 3D perovskite, we mix lead acetate trihydrate ($PbOAc_2$) and methylammonium iodide (MAI) in a molar ratio of 1:3 in DMF at a concentration of 0.7 mol/L. The solution was stirred overnight, and the incorporation of a small amount (<15%) of dimethylammonium cannot be excluded[26]. The solutions were then spin coated on quartz substrates (without antisolvent) and annealed at 150°C for 10 minutes to yield films of 230 nm. Subsequently, the final films are encapsulated with a layer of PMMA.

## III. EXPERIMENTAL RESULTS AND DISCUSSION

### A. Initial Picosecond Response

First, we focus on the early phase of THz-conductivity in 3D and 2D after photoexcitation above bandgap. Figure 2 presents corresponding OPTP (black) and TA traces (photobleach (PB), green) for excitation fluences between 30 to 500 $\mu\frac{\mathrm{J}}{\mathrm{cm}^2}$. The effective carrier densities for MAPI are 3.99, 6.33, 15.88 and $63.25 \cdot 10^{18}\ \mathrm{cm}^{-3}$, as well as 0.82, 1.30, 3.26, 8.20 and $13.00 \cdot 10^{18}\ \mathrm{cm}^{-3}$ for the 2D-PK. The optical transmission increases upon excitation, due to the photobleach, which probes carrier relaxation and population at the conduction band minimum for electrons and valence band maximum for holes, respectively. The THz transmission decreases upon photocarrier generation proportional to the local conductivity, given by the product of the number of carriers and their mobility[27,28]. The full dataset tracks characteristic evolutions and resolves specific differences between 3D and 2D-PK.

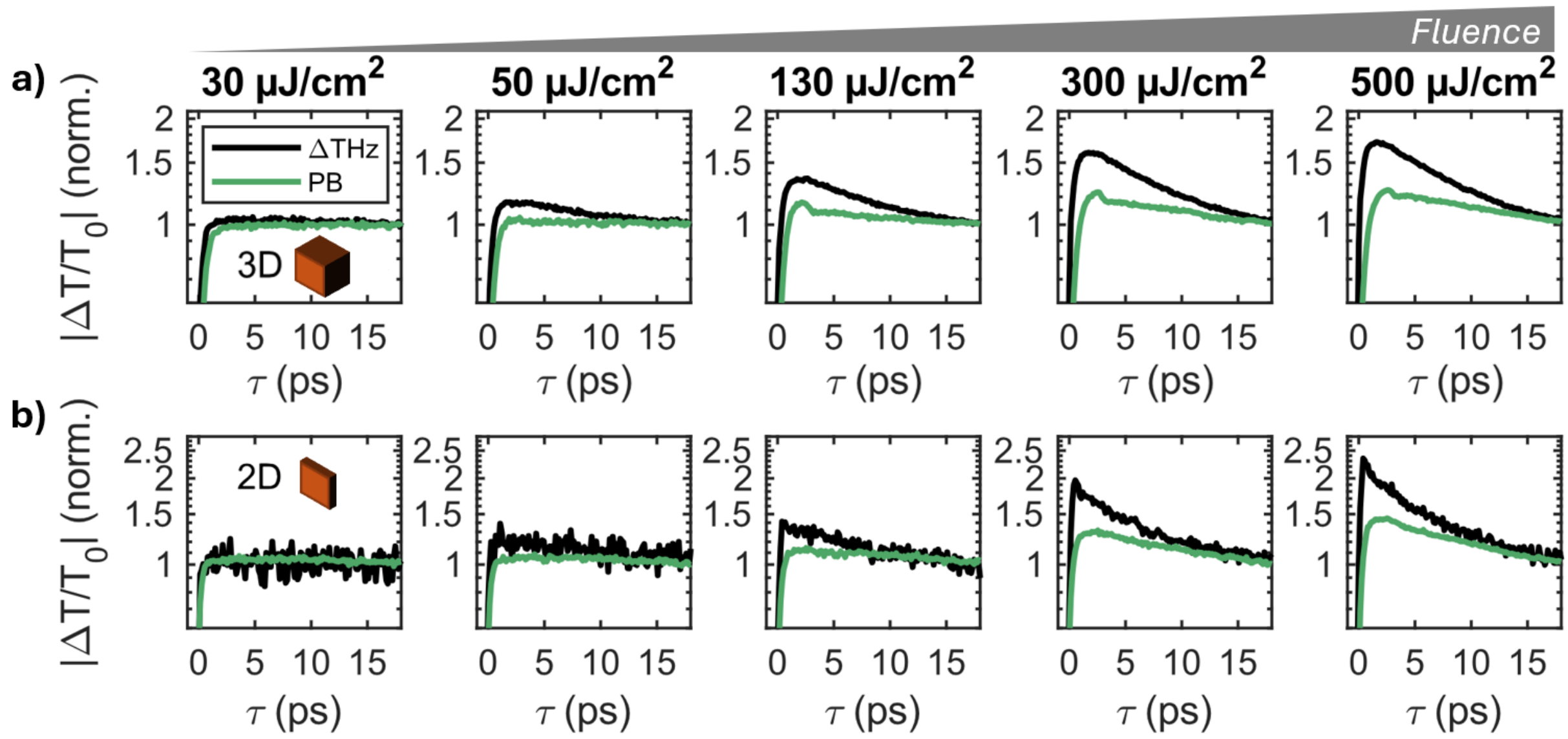

**FIG 2:** Early phases of carrier dynamics (photobleach (PB, green)) and THz-transmission (black) for the 3D and 2D PK. For better comparison, we present absolute (unsigned) values. Excitation fluences increase from 30 to 500 $\mu J/cm^2$. PB probe wavelengths are at 750 nm and 720 nm for the 3D and 2D PK, respectively.

To facilitate the comparison, we now discuss the data in a reduced representation based on two limiting fluences (lowest and highest): As displayed in Fig 3a) for 3D-MAPI, the formation of the full photobleach (PB) at the bandgap shifts for higher excitation towards longer delays (green arrow). In contrast, the formation of the full THz conductivity is not shifting for higher fluence, as indicated by the grey arrow. Figure 3b) illustrates a condensed picture of the subsequent steps of carrier dynamics for 3D: at low intensity (left panel) and above-bandgap photo excitation (1), fast cooling (2) due to efficient energy dissipation via electron-phonon and phonon-phonon coupling results in a quasi-instantaneous photobleach signal caused by the population at the conduction band minimum for electrons and valence band maximum for holes (3). The population decays via radiative and nonradiative recombination (4). In correlation to carrier population, THz radiation probes the carrier conductivity: At low intensity, the THz-absorption essentially follows the PB signal, thus, the conductivity signal is governed by carrier population at the conduction band minimum and valence band maximum for electron / holes, respectively. More precisely, the formation of the full conductivity is reached subtly *before* the PB signals, i.e. within the carrier cooling process. Such behaviour can be related to non-parabolicity in the bands of MAPI, resolved during the cooling process[24].

For higher intensities, the PB data shows a temporal shift towards longer delays that we identify as a hot phonon bottleneck (HPB), as reported for MAPI in various previous studies[12,13,20,14]. The HPB effect prolongs cooling times up to several ps for carrier densities exceeding $10^{18}$ $cm^{-3}$. The slower cooling due to the bottleneck is sketched in Fig. 3b (right panel). Additionally, the hot carrier distributions that contribute to the THz signal during cooling are energetically broad and differences between the slopes of PB and THz signals can be attributed to the admixture of faster decaying higher levels to the THz conductivity, as indicated in Ref.[13].

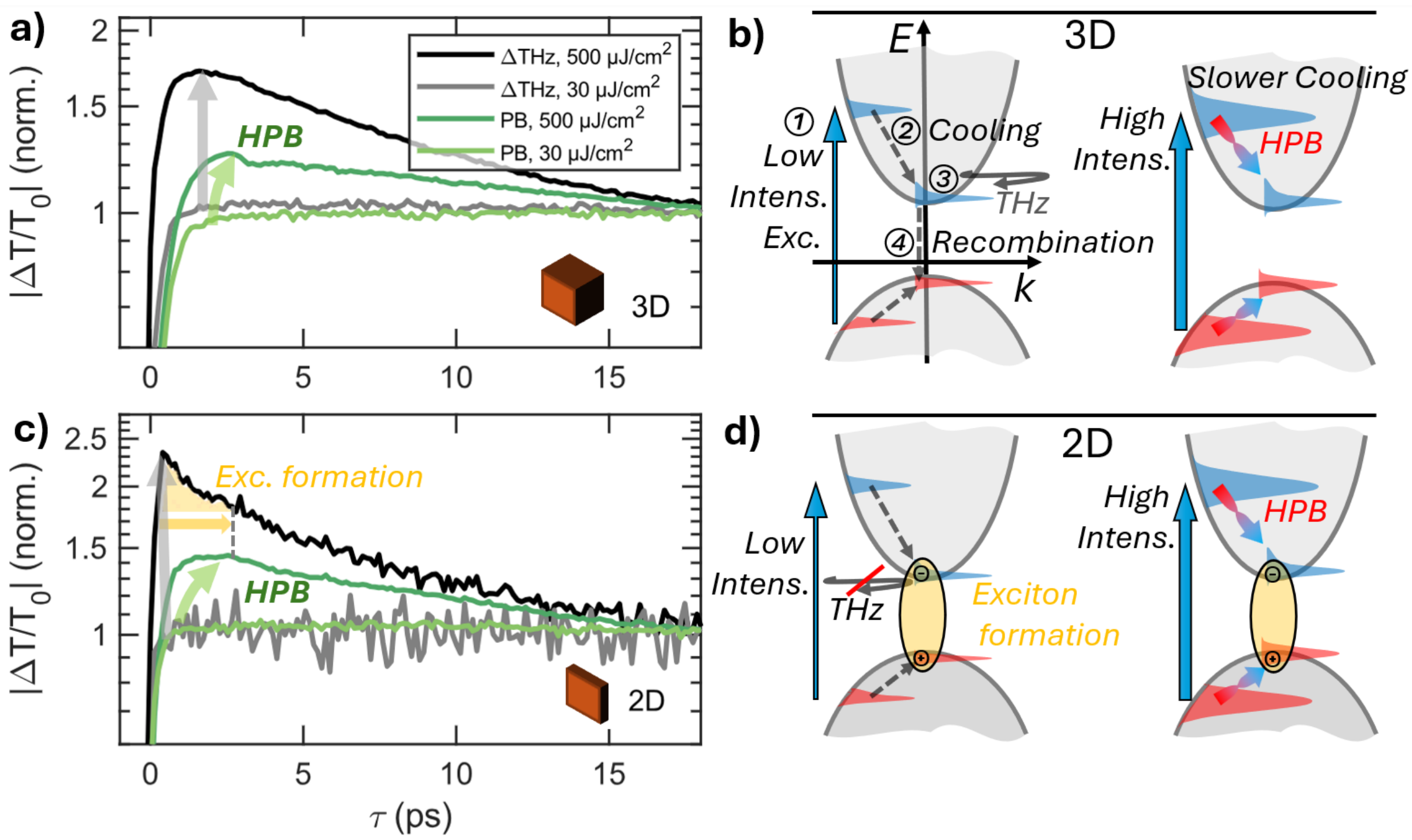

**FIG 3:** Early phases of carrier dynamics and THz-conductivity for 3D and 2D at low/high intensities: a) 3D-MAPI exhibits the full THz conductivity (grey/black) at a picosecond delay without shift for high intensities (grey arrow). In contrast, the photobleach at bandgap (PB, green) shifts to longer delays at higher excitation due to the hot phonon bottleneck (HPB, green arrow). b) Illustration of subsequent steps during carrier dynamics in 3D-PK for two regimes: 1) excitation, 2) cooling, 3) population of band extrema and 4) recombination to ground state. THz probes carrier population and mobility. c) The PB signals in 2D-PK also shifts for increasing intensity. Yet, an additional 2D-specific ultrafast THz-onset and -decay builds-up (shaded yellow). The feature coincides with the hot phonon bottleneck decay in cooling. d) In 2D, photocarriers quasi-instantaneously form excitons due to the enhanced exciton binding energy; such bound electron-hole pairs reduce THz conductivity. At high excitation, however, the reduced cooling (see PB, green arrow in c)) delays the exciton formation and, thus, yields the intensity-dependent picosecond decay of THz-conductivity.

We now compare the 3D-MAPI signals with equivalent data from the 2D-PK in Fig 3c: The key differences are given by a 2D-specific ultrafast onset of THz-conductivity and an additional rapid picosecond decay which builds-up at high intensities. At low intensity, THz data follows the photobleach: We conclude that the conductivity as the product of carrier concentration and mobility is thus governed by the cooled population, revealing that energy-dependencies of mobilities do not significantly impact the evolution of conductivity. For increasing intensity, the PB now exhibits comparable temporal shifts akin to the HPB effect in 3D, consistent with previous results for 2D-PKs with $n \geq 3$. Remarkably, the ultrafast decay feature in the 2D-THz-conductivity coincides temporally with delayed cooling phase in the PB, indicated via the yellow arrow and grey dashed line in Fig. 3c. We attribute this observation to the enhanced exciton binding energy of 100-200 meV, caused by to the dielectric confinement in the lead-iodide-based 2D-PK[1]: Illustrated in Fig. 3d for high intensities, initially hot electrons and holes slowly cool down and form excitons *during* the HPB-delayed cooling process. Such carrier pairs are electrically neutral and exhibit negligible THz absorption outside their resonances. Bound electron-hole pairs are indicated in the single-particle dispersion relation in Fig. 3d via circularized charge carriers (yellow). Thus, the 2D-PK exciton formation generates an efficient additional decay channel for electronic THz-conductivity visible at intense excitation. The fraction of excitons / free electron-holes follows equilibrium dynamics similar to the Saha description in plasma physics, however, significantly modified in the solid state – yielding fractions of several 10% at room temperature, as recently discussed[23,24]. For low intensity, the fast carrier cooling and exciton formation instantaneously reduces the maximum conductivity. For high excitation, however, the slow cooling delays exciton formation and, thus, induces an intensity-dependent ultrafast THz-decay feature. We note that such picosecond decays have been observed in few THz-studies, there, however, independent on intensity and without correlation to the HPB signal in PB[23,24,29] to our knowledge. Our observations are in agreement with recent theoretical analysis of hot carrier cooling in 2D to 3D perovskites[21]: With decreasing confinement from "true" n = 1 2D-PK to 3D, cooling slows down and approaches the bulk value already around n = 3. Correspondingly, the HPB has not been observed for n = 1 2D-PK in Ref.[20]. Thus, the investigated 2D-PK (n = 5) features excitonic properties of a 2D material with carrier cooling related to the bulk-like material. These dynamics potentially have also been implicitly exploited for ultrafast THz photonic devices[30].

### B. Intermediate Sub-Nanosecond Response

Next, we evaluate the intermediate carrier dynamics on longer timescales of several-100 ps. Figure 4 displays correlative time-resolved signals from OPTP, TA and PL for two different excitation fluences. For 3D MAPI, the PB (green) and OPTP (black) signals exhibit essentially identical dynamics after the initial carrier cooling is completed within ~20 ps, as presented in Figs.4a,c. Hence, the decay of THz-conductivity can be attributed predominantly to the dynamics of the carrier population at the conduction band minimum and valence band maximum, respectively. In addition, we include the photoluminescence (tr-PL) trace obtained with a streak-camera/spectrograph combination. The traces were integrated over 30 nm around the peak wavelength, and they are presented as the square root signal of the measured intensity since the bimolecular radiative recombination follows $\propto k_2 n^2$ with carrier density n of a direct bandgap semiconductor. In fact, these overall PL dynamics also mirror the population decay for longer timescales. Differences at early times might be attributed to the impulse response function (~10 ps) and to an excitation wavelength which is blue-shifted compared to the OPTP data, i.e. 415 nm vs. 515 nm.

For the 2D-PK, the complete set of OPTP, BP and PL traces resolve identical dynamics at lower fluence, as shown in Fig.4b. Intriguingly, however, for higher intensity, OPTP and BP reveal different decays at intermediate decays around 200 ps. Such differences might present a hallmark of additional (multi-)excitonic decay channels at high intensities, i.e. bi-excitonic and Auger recombination. Such recombination directly reduces the excitonic parts of population and luminescence, and it only indirectly impacts the free-carrier THz absorption via modifying the exciton / free-carrier equilibrium.

Finally, we present extended sets of OPTP traces in the Supporting Information, and we extracted photophysical rate constants using global fitting *after* the initial carrier cooling phase. We exclude the early phase from the determination of rate constants since the rapid evolution of population and effective carrier temperatures is expected to change the exciton / free-carrier equilibrium and, thus, impact the interpretation of effective rate constants. Similarly to previous reports[29], particularly the fits for slow dynamics caused by carrier trapping (rate constant $k_1$) yield relatively high uncertainties since these are not fully resolved in the data. Still, we observe that the bimolecular "excitonic" recombination (rate constant $k_2$) appears significantly enhanced and dominating for the 2D material - in line recent complimentary spectroscopic results[31]. We note that, in particular for 3D, diffusion along the excitation gradient introduces an additional component to the dynamics which are not yet accounted for in the current model. Future correlative measurements will benefit from the addition of sensitive, spatially resolved measurements[32].

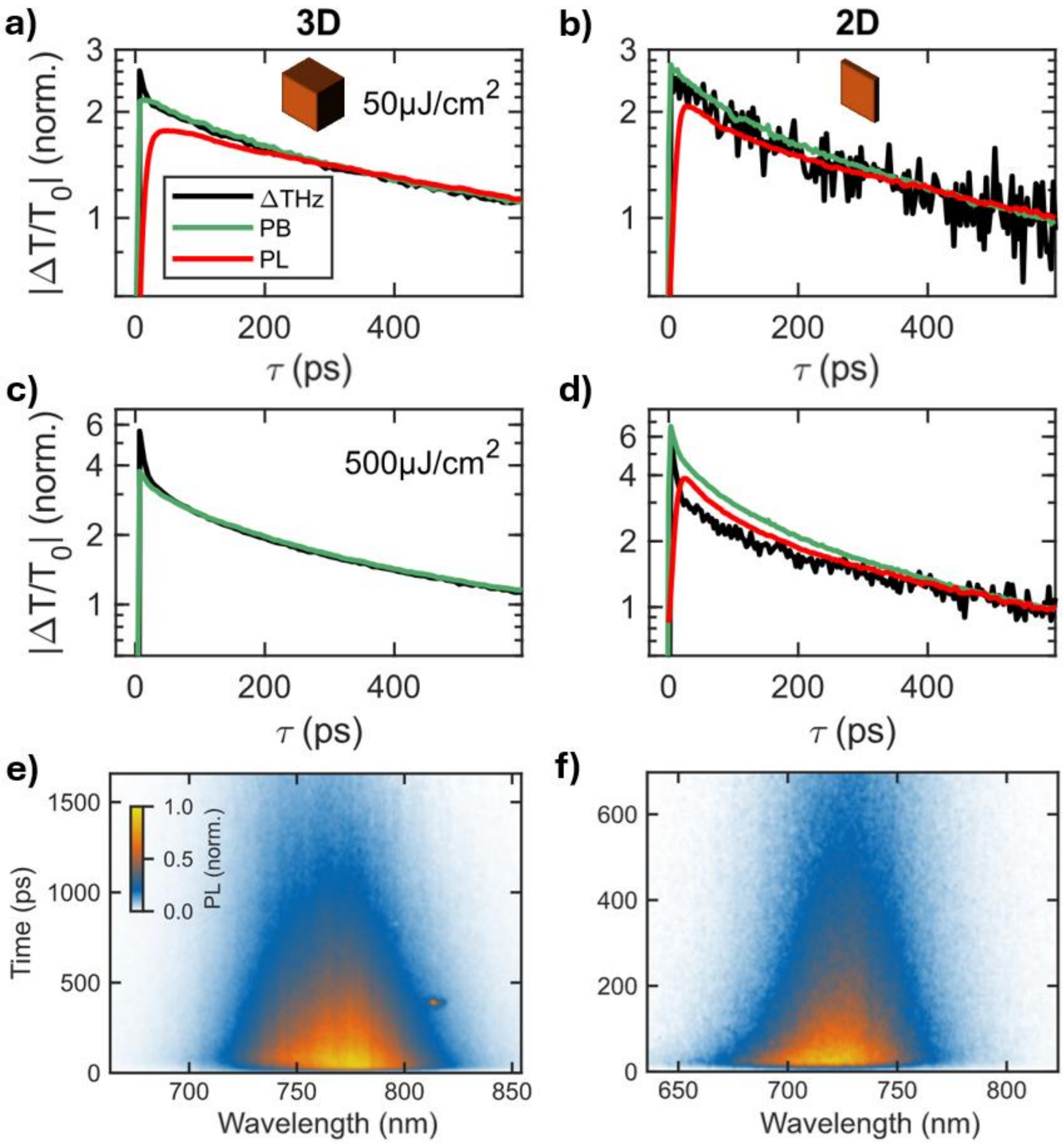

**FIG 4:** Correlative time-resolved signals in the intermediate phase of carrier dynamics for 3D and 2D. a,b) Photobleach (PB, green), THz conductivity (black) and photoluminescence (PL, red)

for the 3D and 2D PK. The PL is included as the square root of the intensity. c,d) Signals at increased intensity. e,f) Characteristic PL streaking traces, recorded at 50 $\mu J/cm^2$.

## IV. CONCLUDING REMARKS

In this study, we investigate the impact of confinement of 3D hybrid perovskite films to 2D on the initial stages of charge carrier generation and transport, using correlative ultrafast spectroscopic techniques covering the NIR and THz spectrum. Combining time-resolved THz spectroscopy with transient absorption and photoluminescence spectroscopy provides access to the dynamics of photo-generated carrier species, populations and mobility in the picosecond to nanosecond range. Specifically, we observe an intensity-dependent picosecond transient in the carrier population of the 2D material that features the signature of a "hot phonon" bottleneck. While this effect has been observed in various 3D hybrid perovskites, the THz conductivity exhibits an accelerated ultrafast 2D-specific feature that has not previously been reported to our knowledge. We attribute the ultrafast conductivity decay to the delayed formation of excitons caused by the blocked cooling of hot carriers. The presented correlative dataset sheds light on the critical interplay between electronic conductivity, hot carrier dynamics with electron-phonon and phonon-phonon interactions, and exciton formation. Further applications of this approach promise to reveal the active role of functional organic interlayers and provide insights into the rational design of tunable, ultrafast carrier responses for future optoelectronic devices.

## ACKNOWLEDGMENTS

We thank J. Köhler for supporting the study with the streak camera setup.

This work was supported by the German Research Foundation (DFG) via CRC 1585 "Structured functional materials for multiple transport in nanoscale confinements", project number 492723217 and projects 403711541, 445471845, 445471097, 445470598.

F.B. acknowledges financial support by the German Research Foundation (DFG) via IRTG 2818 (OPTEXC) and by the Elite Network of Bavaria (study program Biological Physics).

## AUTHOR CONTRIBUTIONS

GH and AJCK devised, planned, and supervised the study. LK, FB designed the OPTP and PB experiments, acquired and processed the data. FB acquired and processed the tr-PL data. LK and FB are co-first authors of this article. MB synthesized the 2D samples, acquired structural data and steady-state optical characterizations. MBH designed the OPTP experiment and acquired tr-PL data. MS, supervised by AK, synthesized the 3D samples and acquired structural data. All authors contributed to the discussion, analysis of the results and writing of the manuscript.